\documentclass{article}
\usepackage{amssymb}
\usepackage{amsmath}
\usepackage[space]{cite}
\usepackage{tikz}
\usepackage{algorithm}
\usepackage{bbm}
\usepackage[noend]{algpseudocode}
\usepackage[normalem]{ulem}
\usepackage{url}
\usepackage{graphicx}
\usepackage{adjustbox}
\usepackage{caption}
\usepackage{subcaption}
\usepackage{enumitem}
\usepackage{xcolor}

\usepackage{MnSymbol}

\usepackage{forest}

\usepackage{cases}

\numberwithin{equation}{section}

\newcommand{\tr}{\mathop{\rm tr}\nolimits}

\newcommand{\atan}{\mathop{\rm atan2}\nolimits}

\usepackage{quantikz}
\usetikzlibrary{circuits.logic.US}
\tikzset{
    gateO/.style={
        draw,
        circle,
        minimum width=0.5em,
        inner sep=2pt    }
}
\DeclareExpandableDocumentCommand{\gateO}{O{}{m}}{|[gateO,#1]| {#2} \qw}

\tikzset{
    gateOS/.style={
        draw,
        circle,
        minimum width=0.5em,
        inner sep=2pt,
		fill=red!20}
}
\DeclareExpandableDocumentCommand{\gateOS}{O{}{m}}{|[gateOS,#1]| {#2} \qw}

\begin{document}

\begin{titlepage}
\vspace{0.5in}

\begin{center}

{\LARGE Preparing multi-qudit states in a definite-weight subspace}\\
\vspace{1in}

\large Nabi Zare Harofteh\footnote{\tt mxz866@miami.edu}\quad and 
\quad Rafael I. Nepomechie\footnote{\tt 
nepomechie@miami.edu}\renewcommand{\thefootnote}{*}\footnote{Corresponding author}
\renewcommand*{\thefootnote}{\arabic{footnote}}
\\[0.2in] 
Department of Physics, PO Box 248046\\[0.2in] 
University of Miami, Coral Gables, FL 33124 USA

\end{center}

\vspace{0.5in}

\begin{abstract}
We formulate a deterministic algorithm for preparing arbitrary multi-qudit states in a definite-weight subspace. By ordering the corresponding computational basis states according to a Gray code for multiset permutations, the state-preparation task is reduced to performing a sequence of controlled 2-qudit Gray rotations. We use this algorithm to prepare exact eigenstates of the $SU(3)$-invariant Heisenberg Hamiltonian, whose Bethe ansatz is nested. In particular, we describe the preparation of the Bethe states, which are $SU(3)$ highest-weight states, 
as well as their lower-weight descendants. We also consider the preparation of $SU(d)$ Dicke states and their $q$-deformations.
\end{abstract}

\end{titlepage}
 
\setcounter{footnote}{0}

\section{Introduction}

The problem of preparing a given multi-qubit quantum state on a quantum computer, using only elementary unitary gates, is an important task in quantum computing \cite{Nielsen:2019}. The problem of preparing exact eigenstates (so-called Bethe states) of the spin-1/2 XXZ chain, which are given by the Bethe ansatz \cite{Bethe:1931hc, Orbach:1958zz, Faddeev:1996iy}, has drawn considerable recent attention \cite{VanDyke:2021kvq, VanDyke:2021nuz, Li:2022czv, Sopena:2022ntq, Ruiz:2023rew, Raveh:2024llj, Ruiz:2025qmt, Sahu:2024toz, Yeo:2025tph}. Since these states are also $U(1)$-eigenstates, they have fixed Hamming weight; hence, algorithms for preparing arbitrary states in a fixed Hamming-weight subspace \cite{Raveh:2024llj, Farias:2024ejh, Mao:2024hfg, Li:2025zzw, Luo:2025zjt}
can in particular be used to prepare such Bethe states. An elegant yet straightforward approach for preparing states in a fixed Hamming-weight subspace, formulated in \cite{Farias:2024ejh}, is to order the corresponding computational basis states
according to a so-called Gray code\footnote{A Gray code due to Ehrlich \cite{Even:1973} is used in \cite{Farias:2024ejh}. For reviews of Gray codes, see e.g. \cite{Wilf:1989, savage:1997, mutze:2012}.}, which reduces the state-preparation task to performing a sequence of controlled 2-qubit rotations.

Since integrable spin-$s$ (where $s \in \{ 1/2, 1, 3/2, \ldots \}$)
generalizations of the XXX model are known \cite{Zamolodchikov:1980ku, Kulish:1981gi, Kulish:1981bi, Babujian:1983ae, Lima:1999, Crampe:2011}, it is natural to consider the problem of preparing the corresponding spin-$s$ Bethe states. These states are also $U(1)$-eigenstates, and therefore have fixed digit sum (rather than
fixed Hamming weight, which is limited to the $s=1/2$ case). An algorithm for preparing arbitrary spin-$s$ $U(1)$-eigenstates (and in particular spin-$s$ Bethe states) was recently formulated in \cite{Harofteh:2026qnh}, based on a Gray code for bounded integer compositions \cite{Walsh:2000}.

The integrable spin-$s$ XXX models are based on $SU(2)$, whose algebra has rank one. Many higher-rank integrable models are also known, see e.g. \cite{Yang:1967bm, Zamolodchikov:1978xm, Jimbo:1985ua, Bazhanov:1986mu}. It is therefore also natural to consider the problem of preparing Bethe states for such higher-rank models, where the so-called nesting phenomenon appears \cite{Yang:1967bm, Sutherland:1975vr, Gaudin:1983}. One of our main goals here is to address this problem for the simplest such model, namely, the $SU(3)$-invariant Heisenberg chain with periodic boundary conditions. Since these Bethe states are linear combinations of computational basis states with definite weights, we first address the more general problem of preparing arbitrary multi-qudit states in a definite-weight subspace, for which the $SU(3)$ Bethe states are a special case. Our approach is again based on a Gray code, here for permutations of a multiset \cite{Ko:1992}, which is sketched in Appendix \ref{sec:KoRuskey}.

The remainder of this paper is organized as follows. In Sec. \ref{sec:basics} we introduce our notations, define the quantum states of interest and introduce the needed Gray code. We present the state-preparation algorithm in Sec. \ref{sec:algorithm}. As simple applications of this result, we consider the preparation of $SU(d)$ Dicke states \cite{Wei:2003, Popkov:2005, Hayashi:2008,  Wei:2008, Zhu:2010, Carrasco:2015sxh} and $SU_q(d)$ Dicke states \cite{Li:2015, Raveh:2023iyy} in Secs. \ref{sec:Dicke} and \ref{sec:qDicke}, respectively. 
However, these states can be prepared more efficiently by exploiting their symmetry.
Our main application, namely the preparation of $SU(3)$ Bethe states, is presented in Sec. \ref{sec:Bethe}; and the preparation of their descendant states is treated in Appendix \ref{sec:descendants}.
We are not aware of alternative proposals for preparing such states. We conclude in Sec. \ref{sec:discussion} with a brief discussion of our results.
The effect of noise on the fidelity of the state-preparation algorithm is briefly considered in Appendix \ref{sec:noise}.
Implementations in cirq \cite{cirq} of the state-preparation algorithm and its applications are available on GitHub \cite{GitHubRN}.

\section{Basics}\label{sec:basics}

We introduce our notations and define the quantum states of interest in Sec. \ref{sec:defs}. The needed Gray code is introduced in Sec. \ref{sec:Gray}.

\subsection{The definite-weight subspace}\label{sec:defs}

We consider $d$-dimensional qudits, using the computational basis states
$|0\rangle, |1\rangle, \ldots |d-1\rangle$ defined as usual
\begin{equation}
|0\rangle=\begin{pmatrix}
1\\
0\\
0\\
\vdots\\
0
\end{pmatrix}\,, \quad| 1\rangle=\begin{pmatrix}
0\\
1\\
0\\
\vdots\\
0
\end{pmatrix} \,,  \ldots \,, \quad
|d-1\rangle =\begin{pmatrix}
0\\
0\\
\vdots\\
0\\
1
\end{pmatrix} \,.
\label{basis}
\end{equation}
We consider $n$-qudit states of the form
\begin{equation}
|\psi(\vec{k})\rangle = \sum_{w \in \mathfrak{S}_{M(\vec k)}} a_w | w \rangle \,,
\label{states}
\end{equation}
where $M(\vec k)$ is the \emph{multiset} (a set with repeated elements) of $n$ integers from 0 to $d-1$
\begin{equation}
M(\vec k)	=\{ \underbrace{0, \ldots, 0}_{k_{0}}, \underbrace{1, 
\ldots, 1}_{k_{1}}, \ldots, \underbrace{d-1, \ldots, d-1}_{k_{d-1}}\} 
= \{ 0^{k_0}, 1^{k_1}, \ldots, (d-1)^{k_{d-1}} \}\,,
\label{multiset}
\end{equation}
where $k_{i}$ is the multiplicity of $i$ in $M(\vec k)$, such that $M(\vec 
k)$ has cardinality $n$. Hence, $\vec k$ is a $d$-dimensional vector such that
\begin{equation}
\vec k = (k_{0}, k_{1}, \ldots, k_{d-1})\quad \text{with}\quad k_{i} \in \{0, 
1, \ldots, n\}\quad \text{and}\quad \sum_{i=0}^{d-1} k_{i} = n\,.
\end{equation}
The sum in \eqref{states} is over the set $\mathfrak{S}_{M(\vec k)}$
of all unique permutations of $M(\vec k)$, whose number $\mathcal{D}$ is given by the multinomial
\begin{equation}
\mathcal{D} = {n \choose \vec k} = {n \choose k_{0}, k_{1}, \ldots, k_{d-1}}	= 
\frac{n!}{\prod_{i=0}^{d-1}k_{i}!} \,.
\label{dim}
\end{equation}
Moreover, $w= w_n \ldots w_1$ with $w_{i} \in \{0, 1, \ldots, d-1\}$ is a permutation in $\mathfrak{S}_{M(\vec k)}$ (we will often refer to $w_1, \ldots, w_n$ as the \emph{components} of 
$w= w_n \ldots w_1$, with corresponding \emph{labels} $1, \ldots, n$), and 
$|w\rangle = |w_n\rangle \otimes \ldots  \otimes |w_1\rangle$
is the corresponding 
$n$-fold tensor product of computational basis states \eqref{basis}. Finally, the coefficients $a_w$ are specified arbitrary complex numbers such that 
$|\psi(\vec k)\rangle$ is normalized 
$\sum_{w \in \mathfrak{S}_{M(\vec k)}} |a_w|^2 = 1$.
An example with $\vec{k}=(2,1,1)$ (and therefore $d=3, n=4$, and $M(\vec{k})=\{0, 0, 1, 2\}$) is
\begin{align}
|\psi(2,1,1)\rangle  &= a_{0012}|0012\rangle  + 
a_{0021}|0021\rangle + a_{0120}|0120\rangle + a_{0102}|0102\rangle
+ a_{0201}|0201\rangle + a_{0210}|0210\rangle \nonumber \\
& + a_{1200}|1200\rangle  + a_{1002}|1002\rangle
+ a_{1020}|1020\rangle + a_{2010}|2010\rangle
+ a_{2001}|2001\rangle  + a_{2100}|2100\rangle \,,
\label{d3example}
\end{align}
where tensor products are suppressed as usual.

Let us define the $U(1)$ generators on a single qudit\footnote{Only $d-1$ of these generators are independent, since $\sum_{i=0}^{d-1}\mathbbm{k}^{(i)} = \mathbb{I}$.}
\begin{equation}
\mathbbm{k}^{(i)} =  |i\rangle\langle i| \,, \qquad i = 0, \ldots, d-1 \,,
\end{equation}
and the corresponding ``total'' $U(1)$ generators (acting
on all $n$ qudits)
\begin{equation}
\mathbb{K}^{(i)} =  \sum_{j=1}^n \mathbbm{k}^{(i)}_j \,, \qquad i = 0, \ldots, d-1 \,,
\label{U1ops}
\end{equation}
which evidently commute $\left[ \mathbb{K}^{(i)} \,, \mathbb{K}^{(i')} \right]=0$.
The states \eqref{states} corresponding to the multiset $M(\vec k)$
are simultaneous eigenstates of all the $\mathbb{K}^{(i)}$ operators, with corresponding eigenvalues $k_i$
\begin{equation}
\mathbb{K}^{(i)} |\psi(\vec{k}) \rangle = k_i\, |\psi(\vec{k}) \rangle\,, \qquad i = 0, \ldots, d-1 \,;
\end{equation}
these states therefore span the subspace with \emph{weight} $\vec{k}$. We will formulate in Sec. \ref{sec:algorithm} an algorithm for preparing such states.

\subsection{Gray code for multiset permutations}\label{sec:Gray}

In order to prepare the states \eqref{states}, we will exploit the remarkable fact (see e.g. \cite{Ko:1992}) that it is always possible to order the permutations $w^{[0]}, w^{[1]}, \ldots \in \mathfrak{S}_{M(\vec k)}$ such that successive permutations $w^{[l]}$ and $w^{[l+1]}$ differ by exactly one interchange (swap) of two distinct elements of the multiset $M(\vec k)$, that is
\begin{equation}
w^{[l+1]} = S_{i^{[l]} j^{[l]}} w^{[l]} \,, \qquad S_{ij} (w_n \ldots w_i \ldots w_j \ldots w_1) = (w_n \ldots w_j \ldots w_i \ldots w_1) \,, \qquad w_i \ne w_j \,,
\label{Grayproperty}
\end{equation}
and every unique permutation appears exactly once. To avoid ambiguity, we define $i, j$ such that $i > j$. An example for the case $\vec{k}=(2,1,1)$ corresponding to the state \eqref{d3example} is displayed in Table \ref{table:KoRuskeyGrayExample}.

\begin{table}[htb]
\centering
\begin{tabular}{|c|c|c|c|c|c|}
\hline
$l$ & $w^{[l]}$ & $i^{[l]}$ & $j^{[l]}$ & $w_{i^{[l]}}^{[l]}$ & $w_{j^{[l]}}^{[l]}$\\   
\hline
0 & 0 0 \textcolor{red}{1} \textcolor{blue}{2} & 2 & 1 & \textcolor{red}{1} &\textcolor{blue}{2}\\
1 & 0 \textcolor{red}{0} 2 \textcolor{blue}{1} & 3 & 1 & \textcolor{red}{0} & \textcolor{blue}{1} \\
2 & 0 1 \textcolor{red}{2} \textcolor{blue}{0} & 2 & 1 & \textcolor{red}{2} & \textcolor{blue}{0} \\
3 & 0 \textcolor{red}{1} 0 \textcolor{blue}{2} & 3 & 1 & \textcolor{red}{1} & \textcolor{blue}{2}\\
4 & 0 2 \textcolor{red}{0} \textcolor{blue}{1} & 2 & 1 & \textcolor{red}{0} & \textcolor{blue}{1}\\
5 & \textcolor{red}{0} 2 \textcolor{blue}{1} 0 & 4 & 2 & \textcolor{red}{0} & \textcolor{blue}{1} \\
6 & 1 \textcolor{red}{2} 0 \textcolor{blue}{0}  & 3 & 1 & \textcolor{red}{2} & \textcolor{blue}{0}\\
7 & 1 0 \textcolor{red}{0} \textcolor{blue}{2} & 2 & 1 &  \textcolor{red}{0} & \textcolor{blue}{2} \\
8 & \textcolor{red}{1} 0 \textcolor{blue}{2} 0  & 4 & 2 & \textcolor{red}{1} & \textcolor{blue}{2}\\
9 & 2 0 \textcolor{red}{1} \textcolor{blue}{0} & 2 & 1 & \textcolor{red}{1} & \textcolor{blue}{0}\\
10 & 2 \textcolor{red}{0} 0 \textcolor{blue}{1} & 3 & 1 & \textcolor{red}{0} & \textcolor{blue}{1} \\
11 & 2 1 0 0 & & & &\\
\hline
\end{tabular}
\caption{A Ko-Ruskey Gray code for $\vec{k}=(2,1,1)$; the $(i^{[l]}\,, j^{[l]})$ values such that $w^{[l+1]} = S_{i^{[l]} j^{[l]}} w^{[l]}$ with $i^{[l]} > j^{[l]}$; and the corresponding values $(w_{i^{[l]}}^{[l]}\,, w_{j^{[l]}}^{[l]})$.}
\label{table:KoRuskeyGrayExample}
\end{table}

We will make use here of an algorithm due to Ko and Ruskey \cite{Ko:1992} for generating permutations of the multiset $M(\vec{k})$ with the 
single-swap ``Gray'' property \eqref{Grayproperty}, which we sketch in Appendix \ref{sec:KoRuskey}. Alternatively, one could search for a
Hamiltonian path on the associated graph, see e.g. \cite{mutze:2012}.

\section{State-preparation algorithm}\label{sec:algorithm}

The state \eqref{states} can now be re-expressed as 
\begin{equation}
|\psi(\vec{k})\rangle = \sum_{l=0}^{\mathcal{D}-1} a_l | w^{[l]} \rangle \,,
\label{states2}
\end{equation}
where $w^{[l]} \in \mathfrak{S}_{M(\vec k)}$ are permutations with the Gray property \eqref{Grayproperty}, and $a_l := a_{w^{[l]}}$.
In order to prepare this state, we first prepare the product state $|w^{[0]} \rangle$, and then (similarly to \cite{Farias:2024ejh, Harofteh:2026qnh}) we perform the successive rotations
\begin{equation}
|w^{[l]} \rangle \rightarrow \cos(\theta_l) | w^{[l]} \rangle + 
e^{i \phi_l} \sin(\theta_l)  | w^{[l+1]} \rangle \,, \quad l=0, 1, \ldots, \mathcal{D}-2 \,,
\label{rotations}
\end{equation}
which results in a state of the form \eqref{states2} with coefficients
\begin{equation}
a_l = \begin{cases}
    \left(\prod_{j=0}^{l-1} e^{i \phi_j} \sin(\theta_j) \right) \cos(\theta_l) & l = 0, \ldots, \mathcal{D}-2 \\[0.1 in]
        \prod_{j=0}^{\mathcal{D}-2}  e^{i \phi_j} \sin(\theta_j) & l = \mathcal{D}-1
    \end{cases} \,.
\label{coeffs}
\end{equation}
The equations \eqref{coeffs} can be inverted to express the required rotation angles
in terms of the specified coefficients $a_0, \ldots, a_{\mathcal{D}-1}$:\footnote{All the coefficients should be rescaled $a_l \mapsto a_l\, |a_0|/a_0$, so that $a_0$ is real.}
\begin{itemize}
\item For the case that all the coefficients are real, all $\phi_l = 0$, and the angles $\theta_l$ are given by 
\begin{equation}
\theta_l = \begin{cases}
\atan \left( \sqrt{\sum_{j=l+1}^{\mathcal{D}-1} a_j^2}, a_l \right) &  l = 0, \ldots, \mathcal{D}-3 \\[0.1 in]
\atan \left(a_{\mathcal{D}-1}, a_{\mathcal{D}-2} \right) & l = \mathcal{D}-2
    \end{cases} \,,
\label{thetas-real}
\end{equation}
where $\atan$ is the 2-argument arctangent function, such that $\atan(y,x)={\rm Arg}(x+i y) \in (-\pi, \pi]$.

\item For the case of complex coefficients, the angles $\theta_l, \phi_l$ are given by
\begin{equation}
\theta_l = \begin{cases}
\atan \left( \sqrt{\sum_{j=l+1}^{\mathcal{D}-1} |a_j|^2}, |a_l| \right) &  l = 0, \ldots, \mathcal{D}-3 \\[0.1 in]
\atan \left(|a_{\mathcal{D}-1}|, |a_{\mathcal{D}-2}| \right) & l = \mathcal{D}-2
    \end{cases} \,,
\label{thetas-complex}
\end{equation}
and
\begin{equation}
\phi_l = 
\arg (a_{l+1}) - \arg  (a_l)\,, \qquad l = 0, \ldots, \mathcal{D}-2 \,.
\label{phis-complex}
\end{equation}

\end{itemize}

\subsection{Gray gate}\label{sec:rotation}

It remains to explain how to implement the $n$-qudit rotation \eqref{rotations}. Due to the Gray property \eqref{Grayproperty}, the states $|w^{[l]}\rangle$ and $|w^{[l+1]}\rangle$ in  \eqref{rotations} are related simply by the interchange of qudits $i^{[l]}$ and $j^{[l]}$ in $|w^{[l]}\rangle$. Hence, the rotation can be accomplished by a transformation, which we call the Gray gate, that performs the corresponding rotation of the relevant pair of qudits.

Specifically, we define the Gray gate $G_{i,j}^{w_i,w_j}(\theta,\phi)$ such that,
when acting on the computational basis states $|\mu \rangle_i\,  |\nu \rangle_j$ with $\mu, \nu  \in \{0, 1, \ldots, d-1\}$ of two qudits labeled by $i, j \in \{1, 2, \ldots, n\}$ with $i > j$, it performs the following unitary transformation  
\begin{equation}
G_{i,j}^{w_i,w_j}(\theta,\phi)\, \left( |\mu \rangle_i\,  |\nu \rangle_j \right) =
\begin{cases}
\cos(\theta)\,  |w_i \rangle_i\,  |w_j \rangle_j
+ e^{i \phi}  \sin(\theta)\,    |w_j \rangle_i\,  |w_i \rangle_j 
& (\mu, \nu) = (w_i, w_j) \\[2pt]
- \sin(\theta)\,  |w_i \rangle_i\,  |w_j \rangle_j
 + e^{i \phi} \cos(\theta)\,  |w_j \rangle_i\,  |w_i \rangle_j
& (\mu, \nu) = (w_j, w_i) \\[2pt]
|\mu \rangle_i\,  |\nu \rangle_j 
& (\mu, \nu) \ne (w_i, w_j)\,, (w_j, w_i)
\end{cases} 
\label{Ggate}
\end{equation}
for given values of $w_i, w_j \in \{0, 1, \ldots, d-1\}$. From the {\em first} line of \eqref{Ggate}, we see that this gate performs (when augmented with suitable controls on other qudits, see Sec. \ref{sec:controls}) the desired $n$-qudit transformation \eqref{rotations}. 

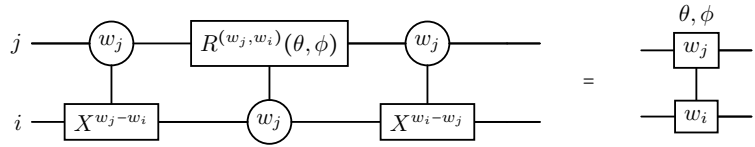
\begin{figure}[htb]
	\centering
\begin{adjustbox}{width=0.6\textwidth}
\begin{quantikz}
\lstick{$j$} & \gateO{w_j} \vqw{1} & \gate{R^{(w_j,w_i)}(\theta,\phi)} \vqw{1} 
& \gateO{w_j} \vqw{1} & \qw & \qw \rstick[2, brackets=none]{$\quad =\quad$}\\
\lstick{$i$} & \gate{X^{w_j - w_i}}  & \gateO{w_j} & \gate{X^{w_i - w_j}}  & \qw & \qw \\
\end{quantikz}
\begin{quantikz}
& \gate[label style={label={above:$\theta, \phi$}}]{w_j} \vqw{1} & \qw \\
& \gate{w_i} & \qw \\
\end{quantikz}
\end{adjustbox}
\caption{Circuit diagram for the Gray gate $G_{i,j}^{w_i,w_j}(\theta,\phi)$ \eqref{Ggate}. In its symbolic form shown on the right, it is understood that the $w$-values in the lower and upper boxes correspond to qudits $i$ and $j$, respectively, as $i > j$.}
\label{fig:Ggate}
\end{figure}

The transformation $G_{i,j}^{w_i,w_j}(\theta,\phi)$ \eqref{Ggate} can be implemented with the quantum circuit shown in Fig. \ref{fig:Ggate}.

\noindent
The horizontal wires represent the two $d$-level qudits.  A circle $\begin{quantikz}\gateO{\scriptscriptstyle \mu}\end{quantikz}$ denotes a control on the value $\mu$. 
The 1-qudit shift gate ${\rm X}$ and its powers are defined as (see e.g. \cite{Wang:2020})
\begin{equation}
{\rm X}\, |\mu\rangle = |\mu+1 \rangle\,, \qquad 
{\rm X}^p\, |\mu\rangle = |\mu+p \rangle \,,
\label{Xgate}
\end{equation}
where the sums are defined modulo $d$; these gates are controlled by the upper wire in Fig. \ref{fig:Ggate}. Moreover, $R^{(w_j,w_i)}(\theta,\phi)$ is the 1-qudit unitary gate 
(Givens rotation) defined by
\begin{align}
R^{(w_j,w_i)}(\theta,\phi) &= 
\cos(\theta) |w_j\rangle\langle w_j | 
-  \sin(\theta) |w_j\rangle\langle w_i |   \nonumber \\
&+ e^{i \phi} \sin(\theta) |w_i\rangle\langle w_j | 
+ e^{i \phi} \cos(\theta) |w_i\rangle\langle w_i | 
+ \sum_{\mu \ne w_i\,, w_j} |\mu\rangle\langle \mu|\,,
\label{Givens}
\end{align}
which in Fig. \ref{fig:Ggate} is controlled by the lower wire. Our Gray gate $G_{i,j}^{w_i,w_j}(\theta,\phi)$ is a qudit generalization of the qubit RBS gate in \cite{Farias:2024ejh}, and is similar to the Gray gate in \cite{Harofteh:2026qnh}.

\subsubsection{Controls on other qudits}\label{sec:controls}

In order to avoid transforming previously-generated states, it is generally necessary to add to the Gray gates $G_{i,j}^{w_i,w_j}$ controls on qudits other than $i, j$. The labels (i.e., addresses, or indices) of these other qudits and the values of the controls can be determined similarly to \cite{Farias:2024ejh, Harofteh:2026qnh}. 
Since all the components of $w^{[l+1]}$ and $w^{[l]}$ are equal except for those labeled $i^{[l]}$ and $j^{[l]}$, the set difference\footnote{By definition, the set difference $A\backslash B$ is the set of elements in set $A$ that are not in set $B$.} 
\begin{equation}
S^{[l]}:=\{1, \ldots, n\}\backslash \{i^{[l]}, j^{[l]}\}
\end{equation}
is the set of labels of components of $w^{[l+1]}$ and $w^{[l]}$ that are {\em equal}. Let $C^{[l]} \subset S^{[l]}$ be the set of labels of components of $w^{[l+1]}$ and $w^{[l]}$ that are {\em equal and nonzero}; that is
\begin{equation}
C^{[l]} := \{ r \in S^{[l]}: w_r^{[l+1]} = w_r^{[l]} \ne 0 \} \,.
\label{setC}
\end{equation}
Naively, a control should be placed at all $r \in C^{[l]}$, with control value $w_r^{[l+1]} = w_r^{[l]}$.

However, some of these controls are redundant, and can be safely removed. The unnecessary controls can be partially pruned using the following
heuristic \cite{Harofteh:2026qnh} that generalizes Subroutine 2 in \cite{Farias:2024ejh}:
Let $U^{[0]}$ be the set of labels of the nonzero ``untouched'' components of the initial permutation $w^{[0]}$; that is, 
\begin{equation}
U^{[0]} := \{ r  \in \{1, \ldots, n\} :\ w^{[0]}_r  \ne 0 \} \,.
\label{setU0}
\end{equation}
At step $l$, define $C^{[l]}$ as before \eqref{setC}; that is
\begin{equation}
C^{[l]} := \{ r \in \{1, \ldots, n\}\backslash \{i^{[l]}, j^{[l]}\}:\ w_r^{[l+1]} = w_r^{[l]} \ne 0 \} \,.
\label{setCagain}
\end{equation}
Set 
\begin{equation}
U^{[l]} =  U^{[l-1]}\backslash \{i^{[l]}, j^{[l]}\} \,, \qquad l>0\,,
\end{equation}
and use it to prune $C^{[l]}$
\begin{equation}
C^{[l]} := C^{[l]}\backslash U^{[l]} \,.
\label{setCpruned}
\end{equation}
Again, a control should be placed at all $r \in C^{[l]}$, with control value $w_r^{[l+1]} = w_r^{[l]}$. These controls should be added to the (single-controlled) Givens 
rotation in Fig. \ref{fig:Ggate}.

\subsection{Summary}\label{sec:full}

We now summarize our algorithm for preparing the state \eqref{states} for a given value of $\vec{k}$ and for specified coefficients $a_{w}$. 

\begin{enumerate}

\item The permutations $w^{[0]}, w^{[1]}, \ldots$ of $M(\vec{k})$ \eqref{multiset} satisfying the Gray property \eqref{Grayproperty} are generated classically, e.g. using the Ko-Ruskey algorithm reviewed in Appendix \ref{sec:KoRuskey}. 

\item The values $(i^{[l]}, j^{[l]})$, $(w_{i^{[l]}}^{[l]}, w_{j^{[l]}}^{[l]})$ and angles $(\theta_l, \phi_l)$ that parametrize the Gray gates \eqref{Ggate}, as well as the labels and values of the controls, are computed classically. The values  $(i^{[l]}, j^{[l]})$, and therefore $(w_{i^{[l]}}^{[l]}, w_{j^{[l]}}^{[l]})$, follow immediately from \eqref{Grayproperty}. The labels and values of the controls are computed from \eqref{setU0}-\eqref{setCpruned}. The angles are given by \eqref{thetas-real} if all the coefficients $a_l := a_{w^{[l]}}$ are real, else by \eqref{thetas-complex} and \eqref{phis-complex}.

\item The quantum part of the algorithm begins by preparing the state $|w^{[0]}\rangle$, which is achieved by applying corresponding powers of
${\rm X}$ gates \eqref{Xgate} to the all-$|0\rangle$ state
\begin{equation}
|w^{[0]}\rangle = \prod_{r=1}^n
{\rm X}^{w^{[0]}_{r}}_r\, |0\rangle^{\otimes n} \,.
\end{equation}

\item Finally, the Gray gates \eqref{Ggate}, controlled according to \eqref{setU0}-\eqref{setCpruned}, are applied successively to prepare the state \eqref{states}
\begin{equation}
|\psi(\vec{k})\rangle = \overset{\curvearrowleft}{\prod_{l=0}^{\mathcal{D}-2}} G^{[l]} \, |w^{[0]}\rangle \,,
\label{genalgorithm}
\end{equation}
where $G^{[l]} = 
G_{i^{[l]},j^{[l]}}^{w^{[l]}_{i^{[l]}},w^{[l]}_{j^{[l]}}}(\theta_l,\phi_l)$,
and the product goes from right to left with increasing $l$. Note that no ancillary qudits are needed.
\end{enumerate}

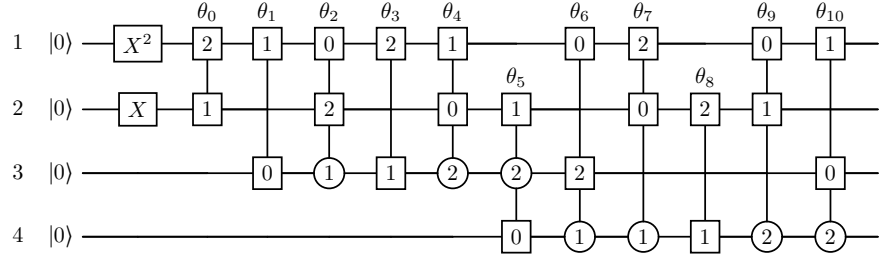
\begin{figure}[htb]
	\centering
\begin{adjustbox}{width=0.7\textwidth}
\begin{quantikz}
\lstick{$1\quad \ket{0}$} & \gate{X^2} 
& \gate[label style={label={above:$\theta_0$}}]{2} \vqw{1} 
& \gate[label style={label={above:$\theta_1$}}]{1} \vqw{2}  
& \gate[label style={label={above:$\theta_2$}}]{0} \vqw{1} 
& \gate[label style={label={above:$\theta_3$}}]{2} \vqw{2}  
& \gate[label style={label={above:$\theta_4$}}]{1} \vqw{1} 
& \qw 
& \gate[label style={label={above:$\theta_6$}}]{0} \vqw{2}
& \gate[label style={label={above:$\theta_7$}}]{2} \vqw{1} 
& \qw
& \gate[label style={label={above:$\theta_9$}}]{0} \vqw{1} 
& \gate[label style={label={above:$\theta_{10}$}}]{1} \vqw{2}
& \qw \\
\lstick{$2\quad \ket{0}$} & \gate{X} 
& \gate{1} & \qw  
& \gate{2} \vqw{1} & \qw 
& \gate{0} \vqw{1}
& \gate[label style={label={above:$\theta_5$}}]{1} \vqw{1} 
& \qw 
& \gate{0} \vqw{2} 
& \gate[label style={label={above:$\theta_8$}}]{2} \vqw{2}  
& \gate{1} \vqw{2} 
& \qw & \qw\\
\lstick{$3\quad \ket{0}$} & \qw  & \qw  & \gate{0} 
& \gateO{1} & \gate{1}  & \gateO{2} & \gateO{2} \vqw{1} 
& \gate{2} \vqw{1}
& \qw  & \qw  & \qw & \gate{0} \vqw{1} & \qw\\
\lstick{$4\quad \ket{0}$} & \qw  & \qw  & \qw  & \qw  & \qw  & \qw 
& \gate{0} & \gateO{1} & \gateO{1} & \gate{1} & \gateO{2} & \gateO{2}
& \qw\\
\end{quantikz}
\end{adjustbox}
\caption{The quantum circuit \eqref{genalgorithm} for preparing a state with $\vec{k} = (2, 1, 1)$, see \eqref{d3example} and Table \ref{table:KoRuskeyGrayExample}. The Gray gates are defined in Fig. \ref{fig:Ggate}, and the circles denote controls.}
\label{fig:CircuitExample}
\end{figure}

For the example \eqref{d3example} with $\vec{k} = (2,1,1)$ that is analyzed in Table \ref{table:KoRuskeyGrayExample}, the quantum circuit \eqref{genalgorithm} is shown in Fig. \ref{fig:CircuitExample}.

The quantum circuit \eqref{genalgorithm} is a sequence of $\mathcal{D}-1$ Gray gates, where $\mathcal{D}$ is given by \eqref{dim}. Each Gray gate \eqref{Ggate} is a 2-qudit Givens rotation dressed with up to $n-2$ additional controls (the set $C^{[l]}$ in Sec. \ref{sec:controls}), which can be synthesized using $\mathcal{O}(n)$ elementary two-qudit gates \cite{Wang:2020, Zi:2023drn}. The total number of two-qudit gates, as well as the circuit depth, is therefore $\mathcal{O}(n \mathcal{D})$. The dominant factor is $\mathcal{D}$;
in the worst case $\vec{k} \sim (\frac{n}{d}\,,\frac{n}{d}\,, \ldots, \frac{n}{d})$, it grows exponentially $\mathcal{D} \sim d^n$,
whereas for many $\vec{k}$ it is only polynomial in $n$.
For example, $\mathcal{D} \sim n^{r x}$
for $\vec{k}$ of the form
\begin{equation}
    \vec{k} = (n - r x, \overbrace{x, \ldots,x}^r, 
    \overbrace{0, \ldots,0}^{d-r-1}) \,,
    \qquad 0 < r\le d-1\,, \qquad 0 < x \le \frac{n}{r+1}\,,
    \label{better}
\end{equation}
(or any permutation thereof) with $x$ constant (independent of $n$). 
More generally, by Stirling, $\ln\mathcal{D}\approx n\ln n-\sum_i k_i\ln
k_i$. If $k^{\text{max}} := \max_i k_i =n-m$
with $m$ fixed (so the remaining $k_i$ contribute $\mathcal{O}(1)$), then
\begin{equation}
\ln\mathcal{D}\approx n\ln n - (n-m)\ln (n-m) \approx m\ln n  \,,
\end{equation}
so $\mathcal{D}\sim n^{m}$ and the gate count is polynomial, as in the
family \eqref{better}. Conversely, $\mathcal{D}$ grows exponentially if $m$ is a fixed fraction of $n$ (equivalently, when at least two of the $k_i$ are proportional to $n$), the extreme case being $\vec k\approx(n/d,\dots,n/d)$, for which $\mathcal{D}\sim d^n$.
We stress that, apart from the quantum cost, there may also be a substantial classical cost for evaluating the coefficients $a_w$, as discussed in Sec. \ref{sec:discussion}.

The circuit requires only single-qudit rotations -- the shift $X$ \eqref{Xgate} and the Givens rotation \eqref{Givens} -- together with two-qudit entangling gates, from which the controlled Gray gates can be compiled. Such a universal gate set has been realized experimentally on both superconducting \cite{Goss:2022, Roy:2022kwg, Nguyen:2023svc} and trapped-ion \cite{Ringbauer:2022, Hrmo:2023} platforms; see also \cite{Wang:2020} for an overview of high-dimensional quantum computing.

Our exact (noiseless) simulations \cite{GitHubRN} validate the algorithm under ideal conditions; we now consider the effect of noise.
Since the circuit is a sequence of $N_g=\mathcal{O}(n\mathcal{D})$ two-qudit gates, under a per-gate incoherent error rate $\epsilon$ the prepared-state fidelity is 
\begin{equation}
F\approx(1-\epsilon)^{N_g}
\approx e^{-\epsilon N_g}\,,  
\label{fidelity}
\end{equation}
see e.g. \cite{Boixo:2016vge}; maintaining a fixed fidelity thus requires a gate error $\epsilon\lesssim 1/N_g$ that shrinks only polynomially with $n$ whenever $\mathcal{D}$ is polynomial. 
We note that the definite-weight structure of the target states permits symmetry-based error detection: since these states are eigenstates of the charges $\mathbb{K}^{(i)}$ \eqref{U1ops}, one may measure the total weight and post-select on the expected value $\vec k$, flagging any error that leaves the weight-$\vec{k}$ subspace. A simple simulation of this protocol is described
in Appendix \ref{sec:noise}. However, a detailed study of decoherence and fault-tolerant implementation on specific hardware is beyond the scope of the present work.

\section{$SU(d)$ Dicke states}\label{sec:Dicke}

A simple class of states \eqref{states} are so-called $SU(d)$ Dicke states $|D^n(\vec{k})\rangle$, defined as uniform superpositions of weight-$\vec{k}$ computational basis states, with coefficients
\begin{equation}
a_w = \frac{1}{\sqrt{\mathcal{D}}} \quad \text{ for all  } 
w \in \mathfrak{S}_{M(\vec k)} \,,
\label{aDicke}
\end{equation}
where $\mathcal{D}$ is given by \eqref{dim}. Returning to the example 
with $\vec{k}=(2,1,1)$, the corresponding $SU(3)$ Dicke state is
\begin{align}
|D^4(2,1,1)\rangle  &= \frac{1}{\sqrt{12}}\Big(|0012\rangle  + 
|0021\rangle + |0120\rangle + |0102\rangle
+ |0201\rangle + |0210\rangle \nonumber \\
& + |1200\rangle + |1002\rangle
+ |1020\rangle + |2010\rangle + |2001\rangle  + |2100\rangle \Big)\,.
\label{Dickeexample}
\end{align}
Such states have been studied over many years, see e.g. \cite{Wei:2003, Popkov:2005, Hayashi:2008,  Wei:2008, Zhu:2010, Carrasco:2015sxh}.

The states $|D^n(\vec{k})\rangle$ are evidently eigenstates of the $U(1)$ operators \eqref{U1ops}
\begin{equation}
\mathbb{K}^{(i)} |D^n(\vec{k}) \rangle = k_i\, |D^n(\vec{k}) \rangle\,, \qquad i = 0, \ldots, d-1 \,.
\end{equation}
These states are also ground states of the $SU(d)$-invariant Hamiltonian
\begin{equation}
H_{SU(d)} = -\sum_{i<j} S_{ij}\,, \qquad S_{ij} = \sum_{a,b=0}^{d-1} e_i^{ab}\, e_j^{ba} \,,
\label{DickeHam}
\end{equation}
where $e^{ab} = |a\rangle \langle b|$ with $a, b = 0, 1, \ldots, d-1$
is the elementary $d \times d$ matrix whose only nonzero matrix element is a 1 at position $(a,b)$, and $e^{ab}_i$ with $i = 1, 2, \ldots, n$
denotes the corresponding matrix at site $i$; hence, $S_{ij}$ is the swap operator for sites $i$ and $j$. The Hamiltonian \eqref{DickeHam} is the quadratic Casimir operator for a chain of $n$ fundamental representations of $SU(d)$, up to an additive constant and an overall factor. The states $|D^n(\vec{k})\rangle$ are eigenstates of this Hamiltonian, with corresponding eigenvalues that are independent of $d$ and $\vec{k}$
\begin{equation}
H_{SU(d)}\, |D^n(\vec{k})\rangle = -\tfrac{1}{2}n(n-1)\, |D^n(\vec{k})\rangle \,.
\label{DickeEig}
\end{equation}

We have explicitly verified that, using the coefficients \eqref{aDicke}, the circuit \eqref{genalgorithm} indeed prepares the exact states $|D^n(\vec{k})\rangle$, and we have explicitly verified the property \eqref{DickeEig},
for various values of $\vec{k}$, see \cite{GitHubRN}. The high degree of symmetry of these states can be exploited to prepare these states more efficiently \cite{Nepomechie:2023lge, Raveh:2024sku, Liu:2024taj, Kerzner:2025uxw}.

\section{$SU_q(d)$ Dicke states}\label{sec:qDicke}

Another simple class of states \eqref{states} are $SU_q(d)$ Dicke states $|D^n_q(\vec{k})\rangle$, whose coefficients are given for $q>0$ by \cite{Raveh:2023iyy}
\begin{equation}
a_w = \frac{1}{\sqrt{{n \brack \vec k}}} q^{\frac{J(\vec k)}{2}-\text{inv}(w)} \,,
\label{aqDicke}
\end{equation}
where $\text{inv}(w)$ denotes the inversion number (the minimum number of adjacent transpositions needed to go from the identity permutation to $w$) of the permutation $w$,
and $J(\vec k) =\sum_{i<j} k_i\, k_j$ is the maximum inversion number. Moreover, 
${n \brack \vec k}$ denotes the $q$-multinomial
\begin{equation}
{n \brack \vec k}={n \brack k_0,k_1,\dots,k_{d-1}}=\frac{[n]!}{\prod_{i=0}^{d-1}[k_i]!} \,,
\label{qmultinomial}
\end{equation}
the $q$-factorial is defined for
non-negative integers $n$ by
\begin{equation}
[n]!=[n][n-1]\dots[1],\quad\text{with}\quad[0]!=1 \,,
\end{equation}
and 
\begin{equation}
[x]=\frac{q^x-q^{-x}}{q-q^{-1}}\,,
\label{bracket}
\end{equation}
so that $[x]$ is symmetric under $q\to1/q$. It is clear that $[x]\to x$ as $q\to1$, so that $[x]$ is a $q$-deformation of $x$. 
Hence, the states $|D_q^n(\vec{k})\rangle$ reduce to the $SU(d)$ Dicke states $|D^n(\vec{k})\rangle$ in the limit $q \rightarrow 1$. The case $d=2$ was first considered in \cite{Li:2015}.
For the example $\vec{k}=(2,1,1)$, the corresponding $SU_q(3)$ Dicke state is
\begin{align}
|D^4_q(2,1,1)\rangle   &= \frac{1}{\sqrt{q^5+2q^3+3q+3q^{-1}+2q^{-3}+q^{-5}}}
\Big(q^{\frac{5}{2}}|0012\rangle  + 
q^{\frac{3}{2}}|0021\rangle \nonumber \\
& + q^{\frac{1}{2}}|0120\rangle + q^{\frac{3}{2}}|0102\rangle
+ q^{\frac{1}{2}}|0201\rangle + q^{-\frac{1}{2}}|0210\rangle 
 + q^{-\frac{3}{2}}|1200\rangle  \nonumber \\
& + q^{\frac{1}{2}}|1002\rangle
+ q^{-\frac{1}{2}}|1020\rangle  
+ q^{-\frac{3}{2}}|2010\rangle
+ q^{-\frac{1}{2}}|2001\rangle  
+ q^{-\frac{5}{2}}|2100\rangle \Big)\,.
\label{qDickeexample}
\end{align}
Indeed, recall from \eqref{aqDicke} that the coefficient of $|w\rangle$ is proportional to $q^{J(\vec{k})/2-\text{inv}(w)}$, where
$\text{inv}(w)$ is the number of inversions of the string $w=w_n\cdots
w_1$ (the number of pairs of sites whose values are out of ascending
order), and $J(\vec{k})$ is the maximal inversion number, attained by
the fully reversed string. For $\vec k=(2,1,1)$, we have
$J(\vec{k})=\text{inv}(2100)=5$ (equivalently, 
$J(\vec{k})=\sum_{i<j} k_i\, k_j = 2\cdot 1+2\cdot 1+1\cdot 1=5$).
Hence, for example, $\text{inv}(0012)=0$ gives the coefficient 
$q^{\frac{5}{2}-0}=q^{\frac{5}{2}}$, while $\text{inv}(0021)=1$ gives 
$q^{\frac{5}{2}-1}=q^{\frac{3}{2}}$, in agreement with \eqref{qDickeexample}.
For the case of $q$ complex, the coefficients are again given by \eqref{aqDicke}, except that the normalization factor is computed with $q$ replaced by $|q|$ \cite{Raveh:2023iyy}.

In addition to being eigenstates of the $U(1)$ operators \eqref{U1ops}
\begin{equation}
\mathbb{K}^{(i)} |D_q^n(\vec{k}) \rangle = k_i\, |D_q^n(\vec{k}) \rangle\,, \qquad i = 0, \ldots, d-1 \,,
\end{equation}
the states $|D_q^n(\vec{k})\rangle$ can be shown to be eigenstates of a $q$-deformation of the Hamiltonian \eqref{DickeHam}. Indeed, define
\begin{equation}
H_{SU_q(d)} = -\sum_{i=2}^n J_i \,, \qquad J_i = \check{R}_{i-1} + \check{R}_{i-1}\, J_{i-1}\, \check{R}_{i-1} \,, \qquad J_1 = 0\,, 
\label{qDickeHam}
\end{equation}
where $J_i$ are $q$-deformations of so-called Jucys-Murphy elements (see e.g. \cite{Katriel:1995, Doikou:2009} and references therein), which 
have the commutativity property
\begin{equation}
    \left[ J_i \,, J_j \right] = 0 \,,
\end{equation}
and are expressed in terms of the Hecke operators
\begin{equation}
\check{R} = q \sum_a e^{aa} \otimes e^{aa} + \sum_{a\ne b} e^{ab} \otimes e^{ba} + (q-q^{-1})\sum_{a < b} e^{aa} \otimes e^{bb} \,.
\end{equation}
Note that the Hecke operators, which satisfy the relations
\begin{align}
\check{R}_i\, \check{R}_{i+1}\, \check{R}_i &=
\check{R}_{i+1}\, \check{R}_i\, \check{R}_{i+1}\,, \nonumber \\
\check{R}_i\, \check{R}_j &= \check{R}_j\, \check{R}_i \,, \qquad |i-j| \ge 2 \,, \nonumber \\ 
(\check{R}_i - q) (\check{R}_i + q^{-1}) &= 0 \,,
\end{align}
act nontrivially on two sites; $\check{R}_i$ acts nontrivially on sites $i$ and $i+1$. The Hamiltonian \eqref{qDickeHam} has only $U(1)^{d-1}$ (rather than $SU(d)$) symmetry.
In the limit $q \rightarrow 1$, the Hamiltonian 
\eqref{qDickeHam} reduces to \eqref{DickeHam}.
The states $|D_q^n(\vec{k})\rangle$ are eigenstates of the Hamiltonian \eqref{qDickeHam}
\begin{equation}
H_{SU_q(d)}\, |D_q^n(\vec{k})\rangle = E\, |D_q^n(\vec{k})\rangle \,,
\qquad E = -\sum_{i=0}^{n-2} (i+1) q^{2n-3-2i} \,,
\label{qDickeEig}
\end{equation}
which does not seem to have been previously noticed in the physics literature.

We have explicitly verified that, using the coefficients \eqref{aqDicke},
the circuit \eqref{genalgorithm} indeed prepares the exact states $|D_q^n(\vec{k})\rangle$, and we have explicitly verified the property \eqref{qDickeEig}, for various values of $\vec{k}$ and $q$, see \cite{GitHubRN}. The high degree of symmetry of these states can be exploited to prepare these states more efficiently \cite{Raveh:2023iyy}.

\section{$SU(3)$ Bethe states}\label{sec:Bethe}

The $SU(2)$-invariant Heisenberg Hamiltonian with periodic boundary conditions, whose integrability was revealed by Bethe \cite{Bethe:1931hc}, has a natural generalization to $SU(d)$ 
\begin{equation}
H= \sum_{i=1}^n \left(S_{i,i+1} - \mathbb{I}\right) \,, \qquad S_{n,n+1} = S_{n,1} \,,
\label{HeisHam}
\end{equation}
where $S_{ij}$ is the swap operator for $d$-level qudits
defined in \eqref{DickeHam}. Remarkably, this $SU(d)$-invariant Hamiltonian is integrable for all integer values $d \ge 2$ \cite{Sutherland:1975vr, Gaudin:1983}. However, for $d>2$, the Bethe ansatz solution entails an intricate nesting procedure. For simplicity, we focus here on the case $d=3$, where the nesting phenomenon first appears.

An expression for the $SU(3)$ Bethe states (exact eigenstates of the Hamiltonian \eqref{HeisHam} with $d=3$ in terms of two sets of Bethe roots $\{u_1, \ldots, u_{m_1} \}$ and $\{v_1, \ldots, v_{m_2}\}$ with $0\le m_2 \le m_1 \le n$, known as coordinate Bethe ansatz) is given by \cite{Yang:1967bm, Sutherland:1975vr, Gaudin:1983, Ovchinnikov:2010vb}
\begin{equation}
|\{u_1, \ldots, u_{m_1} \}; \{v_1, \ldots, v_{m_2} \}
\rangle = \sum_{1 \le x_1 < \ldots < x_{m_1} \le n}\
\sideset{}{'}\sum_{\alpha_1, \ldots, \alpha_{m_1} = 1}^{2}   f\overset{(\alpha_1, \ldots, \alpha_{m_1})}{(x_1, \ldots, x_{m_1}}; \{ u\}, \{v\})\, e_{x_1}^{\alpha_1,0} \cdots e_{x_{m_1}}^{\alpha_{m_1},0} |0\rangle^{\otimes n} \,,
\label{BA1}
\end{equation}
where the prime on the summation over $\alpha$'s denotes the restriction
\begin{equation}
\sum_{j=1}^{m_1} \delta_{2, \alpha_j} = m_2  \,;
\end{equation}
that is, the summation is over all $\alpha_j \in \{1, 2\}$ such that 
the number of $\alpha$'s with value 2 is $m_2$, and the $m_1-m_2$ remaining $\alpha$'s have value 1. Moreover, the matrices $e^{ab}_i$ in \eqref{BA1} are the same as in \eqref{DickeHam}; hence, $e^{10} |0\rangle = |1\rangle$ and $e^{20} |0\rangle = |2\rangle$.
We define $y_1, \ldots, y_{m_2}$ with $y_l \in \{1, \ldots, m_1\}$
as the positions of those $\alpha$'s with value 2; that is,
\begin{equation}
\alpha_{y_l}=2 \,, \qquad l = 1, \ldots, m_2 \,.
\label{ydef}
\end{equation}
The function $f$ in \eqref{BA1} is given by
\begin{align} 
f\overset{(\alpha_1, \ldots, \alpha_{m_1})}{(x_1, \ldots, x_{m_1}}; \{ u\}, \{v\}) &= \sum_{P \in \mathfrak{S}_{m_1}} \sum_{Q \in \mathfrak{S}_{m_2}} \Big\{ \varepsilon(P) \varepsilon(Q) A(u_{P(1)}, \ldots, u_{P(m_1)})\, B(v_{Q(1)}, \ldots, v_{Q(m_2)}) \nonumber  \\
& \times \prod_{j=1}^{m_1}
\left(\frac{u_{P(j)}+\frac{i}{2}}{u_{P(j)}-\frac{i}{2}}\right)^{x_j}
\prod_{l=1}^{m_2} \Phi(v_{Q(l)}; u_{P(1)}, \ldots, u_{P(y_l)}) \Big\} \,,
\label{BA2}
\end{align}
where the sums are over the sets $\mathfrak{S}_{m_1}$ and $\mathfrak{S}_{m_2}$ of all permutations of $\{1, \ldots, m_1\}$ and $\{1, \ldots, m_2\}$, respectively; and $\varepsilon=\pm 1$ denotes the sign of a permutation. The functions $A, B, \Phi$ in \eqref{BA2} are given by
\begin{align}
A(u_1, \ldots, u_{m_1}) &= \prod_{1 \le a < b \le m_1} (u_a - u_b +i) \,, \nonumber\\
B(v_1, \ldots, v_{m_2}) &= \prod_{1 \le a < b \le m_2} (v_a - v_b +i) \,, \nonumber\\
\Phi(v; u_1, \ldots, u_y) &= \frac{1}{v-u_y-\frac{i}{2}} 
\prod_{l=1}^{y-1}\frac{v-u_l+\frac{i}{2}}{v-u_l-\frac{i}{2}} \,.
\label{BA3}
\end{align}
The Bethe roots $\{u_1, \ldots, u_{m_1} \}$ and $\{v_1, \ldots, v_{m_2}\}$ are solutions of the Bethe equations
\begin{align}
\left(\frac{{u_j}-\frac{i}{2}}{{u_j}+\frac{i}{2}}\right)^n &=
\prod_{l \ne j}^{m_1} \frac{u_j - u_l -i}{u_j - u_l +i} 
\prod_{l =1}^{m_2} \frac{u_j - v_l +\frac{i}{2}}{u_j - v_l -\frac{i}{2}} \,, \qquad
j = 1, \ldots, m_1 \,, \label{BE1} \\
 1 &=
\prod_{l=1}^{m_1} \frac{v_j - u_l + \frac{i}{2}}{v_j - u_l - \frac{i}{2}}
\prod_{l\ne j}^{m_2} \frac{v_j - v_l -i }{v_j - v_l +i} \,, \qquad
j = 1, \ldots, m_2 \,.
\label{BE2}
\end{align}
Various approaches have been developed for solving such equations.
For small values of $n$, a particularly useful approach is to exploit the corresponding Q-system \cite{Marboe:2016yyn}.

The Bethe states \eqref{BA1} are exact eigenstates of the $d=3$ Hamiltonian \eqref{HeisHam}
\begin{equation}
H\, |\{u \}; \{v \}
\rangle = E\, |\{u \}; \{v \}
\rangle \,, 
\end{equation}
with corresponding energy
\begin{equation}
E = -\sum_{j=1}^{m_1} \frac{1}{u_j^2 + \frac{1}{4}} \,,
\label{energyBA}
\end{equation}
as well as of the $U(1)$ operators \eqref{U1ops}
\begin{equation}
\mathbb{K}^{(i)} |\{u \}; \{v \}
\rangle = k_i\, |\{u \}; \{v \}
\rangle\,, \qquad i = 0, 1, 2 \,,
\end{equation}
where
\begin{equation}
k_0 = n - m_1 \,, \qquad k_1 = m_1 - m_2 \,, \qquad k_2 = m_2 \,.
\label{kBA}  
\end{equation}

The Bethe state \eqref{BA1} can be cast in the form \eqref{states}, with $\vec{k}$ given by \eqref{kBA}, and with coefficients $a_w$ given (up to an overall normalization factor) by
\begin{equation}
a_{w_n \ldots w_1} = f\overset{(\alpha_1, \ldots, \alpha_{m_1})}{(x_1, \ldots, x_{m_1}}; \{ u\}, \{v\}) \,,
\label{awBA}
\end{equation}
where the $x_i$'s are the positions of the nonzero $w_i$'s, and the 
$\alpha$'s are the values of those nonzero $w_i$'s.

Using Bethe roots\footnote{For simplicity, we restrict to regular (non-singular) Bethe roots.} obtained using the $SU(3)$ Q-system \cite{Marboe:2016yyn},
we have explicitly verified that the circuit \eqref{genalgorithm} indeed prepares exact eigenstates of the Hamiltonian \eqref{HeisHam}, with corresponding eigenvalue \eqref{energyBA}, for various values of $n, m_1, m_2$, see \cite{GitHubRN}.

The Bethe states \eqref{BA1} are $SU(3)$ highest-weight states \cite{Liashyk:2026dhm}. The lower-weight descendant states can be prepared similarly to the Bethe states, except with different amplitudes, as shown in Appendix \ref{sec:descendants}.

We expect that $SU(d)$ Bethe states with $d>3$ can be prepared by the same circuit construction; the only new ingredient is the corresponding coordinate Bethe wavefunction, which entails $(d-1)$-fold nesting and becomes increasingly intricate -- beyond \eqref{BA2}-\eqref{BA3} -- as $d$ grows. We leave the elucidation of these higher-rank wavefunctions to future work.

\section{Discussion}\label{sec:discussion}

We have formulated an algorithm for deterministically preparing arbitrary multi-qudit states in a weight-$\vec{k}$ subspace \eqref{states}. This algorithm is similar to those for preparing multi-qubit states in a fixed Hamming-weight subspace \cite{Farias:2024ejh} and spin-$s$ $U(1)$ eigenstates \cite{Harofteh:2026qnh}, in the sense that all three algorithms exploit Gray codes to reduce the state-preparation task to performing a sequence of controlled 2-qudit Gray rotations. However, these algorithms all require different Gray codes, and therefore they differ in crucial details. We cannot help but marvel at the fact that the needed Gray codes \cite{Even:1973, Walsh:2000, Ko:1992} were discovered decades earlier in the mathematics/computer science communities.

We have demonstrated that this algorithm can in principle be used to prepare nested Bethe states, which therefore opens the way for preparing Bethe states for many classes of integrable models. Indeed, in our implementation of the algorithm using the cirq statevector simulator \cite{cirq}, one simply inputs the system size and Bethe roots, and the output is the corresponding explicit Bethe statevector \cite{GitHubRN}.
However, as emphasized in \cite{Raveh:2024llj, Harofteh:2026qnh}, this 
coordinate-Bethe-ansatz approach has a high classical overhead for computing the coefficients $a_w$, and therefore for computing the corresponding angles of the Gray rotations. Moreover, the quantum complexity of the algorithm is also high.
Hence, this approach is not practical for large system size.

We note that relatively few explicit expressions for nested coordinate Bethe ansatz wavefunctions are available in the literature, presumably partly because of their unwieldiness, as well as due to the advent of nested algebraic Bethe ansatz \cite{Kulish:1979cr}. Indeed, we did not succeed to find anywhere in the literature
the explicit expression \eqref{BA2}-\eqref{BA3} for the $SU(3)$ wavefunction, which we had to reconstruct from \cite{Yang:1967bm, Sutherland:1975vr, Gaudin:1983, Ovchinnikov:2010vb}. Hence, we expect that new nested coordinate Bethe ansatz
results may be needed in order to prepare in this way eigenstates of other integrable models, such as those with anisotropy, boundaries, or yet-higher rank. For the case of boundaries, our algorithm is applicable provided that the boundary interactions do not break the 
$U(1)^{d-1}$ bulk symmetries.

It remains to be seen whether an alternative Bethe-state-preparation approach based instead on algebraic Bethe ansatz \cite{Faddeev:1996iy, Sopena:2022ntq, Ruiz:2023rew, Ruiz:2025qmt, Kulish:1979cr} can yield similarly explicit results while overcoming the challenges of high classical and quantum complexity.

\section*{Acknowledgements}
We thank Eric Ragoucy for valuable correspondence, and for sharing with us the unpublished manuscript \cite{Alonzi:2010}. Generative AI tools (Gemini 3.1 and Opus 4.8) were used for finalizing the nested coordinate Bethe ansatz expressions \eqref{BA2}-\eqref{BA3}, determining the amplitudes for descendant states (Sec. \ref{sec:descendants}), supporting literature review, and for generating some of the figures and code. RN is supported in part by the National Science Foundation under grants PHY 2310594 and PHY 2609869, and by a Cooper fellowship.  

\appendix

\section{A Ko-Ruskey Gray code for multiset permutations}\label{sec:KoRuskey}

\begin{figure}[htbp]
    \centering
    \begin{adjustbox}{width=0.5\textwidth}
    \begin{forest}
      for tree={
        circle,
        draw,
        inner sep=1pt,
        minimum size=12pt,
        s sep=1mm,   
        l sep=8mm,  
        font=\sffamily\footnotesize
      }
      [root, draw=none, rectangle, font=\bfseries
        [0
          [0
            [1 [2 [0012, no edge, draw=none, rectangle, font=\bfseries\footnotesize]]]
            [2 [1 [0021, no edge, draw=none, rectangle, font=\bfseries\footnotesize]]]
          ]
          [1
            [0 [2 [0102, no edge, draw=none, rectangle, font=\bfseries\footnotesize]]]
            [2 [0 [0120, no edge, draw=none, rectangle, font=\bfseries\footnotesize]]]
          ]
          [2
            [0 [1 [0201, no edge, draw=none, rectangle, font=\bfseries\footnotesize]]]
            [1 [0 [0210, no edge, draw=none, rectangle, font=\bfseries\footnotesize]]]
          ]
        ]
        [1
          [0
            [0 [2 [1002, no edge, draw=none, rectangle, font=\bfseries\footnotesize]]]
            [2 [0 [1020, no edge, draw=none, rectangle, font=\bfseries\footnotesize]]]
          ]
          [2
            [0 [0 [1200, no edge, draw=none, rectangle, font=\bfseries\footnotesize]]]
          ]
        ]
        [2
          [0
            [0 [1 [2001, no edge, draw=none, rectangle, font=\bfseries\footnotesize]]]
            [1 [0 [2010, no edge, draw=none, rectangle, font=\bfseries\footnotesize]]]
          ]
          [1
            [0 [0 [2100, no edge, draw=none, rectangle, font=\bfseries\footnotesize]]]
          ]
        ]
      ]
    \end{forest}
    \end{adjustbox}
    \caption{Tree for lexicographic ordering of multiset permutations for $\vec{k}=(2,1,1)$, which does \emph{not} have the Gray property.}
    \label{fig:naive}
\end{figure}
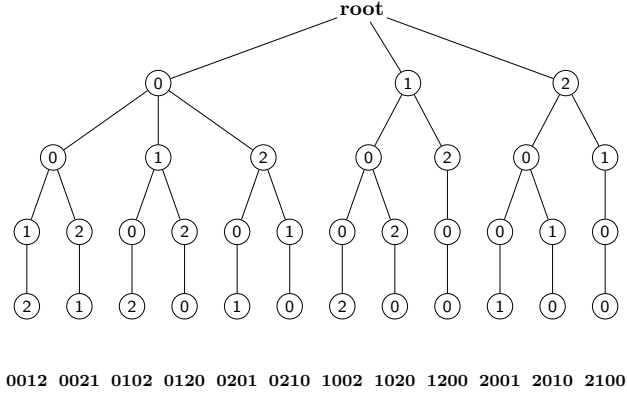

It is convenient to represent the permutations of a multiset $M(\vec{k})$ by an upside-down tree, as shown in Fig. \ref{fig:naive} for $\vec{k}=(2,1,1)$, such that every path from the root down to a leaf spells out one of the permutations. In the naive recursive scheme shown in this figure (corresponding to the \texttt{GenBag} algorithm in \cite{Ko:1992}), the children (the remaining available elements of the multiset)
at each node are selected in strictly increasing order, resulting in a standard lexicographic (dictionary) ordering of the permutations (reading the bold permutations across the bottom from left to right), which does \emph{not} satisfy the single-swap (Gray) property \eqref{Grayproperty}.

\begin{figure}[htbp]
    \centering
    \begin{adjustbox}{width=0.5\textwidth}
    \begin{forest}
      for tree={
        circle,
        draw,
        inner sep=1pt,
        minimum size=12pt,
        s sep=1mm,   
        l sep=8mm,  
        font=\sffamily\footnotesize
      }
      [root, draw=none, rectangle, font=\bfseries
        [0+
          [0+
            [1+ [2 [0012, no edge, draw=none, rectangle, font=\bfseries\footnotesize]]]
            [2-- [1 [0021, no edge, draw=none, rectangle, font=\bfseries\footnotesize]]]
          ]
          [1--
            [2+ [0 [0120, no edge, draw=none, rectangle, font=\bfseries\footnotesize]]]
            [0-- [2 [0102, no edge, draw=none, rectangle, font=\bfseries\footnotesize]]]
          ]
          [2+
            [0+ [1 [0201, no edge, draw=none, rectangle, font=\bfseries\footnotesize]]]
            [1-- [0 [0210, no edge, draw=none, rectangle, font=\bfseries\footnotesize]]]
          ]
        ]
        [1--
          [2--
            [0+ [0 [1200, no edge, draw=none, rectangle, font=\bfseries\footnotesize]]]
          ]
          [0+
            [0-- [2 [1002, no edge, draw=none, rectangle, font=\bfseries\footnotesize]]]
            [2+ [0 [1020, no edge, draw=none, rectangle, font=\bfseries\footnotesize]]]
          ]
        ]
        [2+
          [0--
            [1-- [0 [2010, no edge, draw=none, rectangle, font=\bfseries\footnotesize]]]
            [0+ [1 [2001, no edge, draw=none, rectangle, font=\bfseries\footnotesize]]]
          ]
          [1+
            [0-- [0 [2100, no edge, draw=none, rectangle, font=\bfseries\footnotesize]]]
          ]
        ]
      ]
    \end{forest}
    \end{adjustbox}
    \caption{Tree for multiset permutations with the Gray property
    for $\vec{k}=(2,1,1)$. Note that the signs of the nodes alternate across each level. (Adapted from Fig. 3 in \cite{Ko:1992}.)}
    \label{fig:GenAltA}
\end{figure}
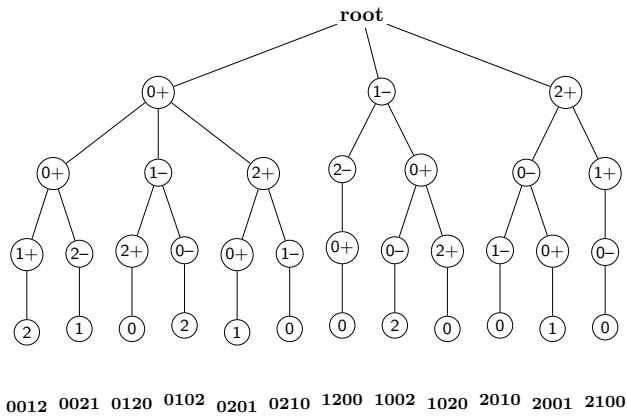

An ingenious way to ensure the single-swap property between successive permutations is to assign signs $(\pm)$ to the nodes: the children at a plus-node $(+)$ are selected in increasing order, while the children at a negative-node $(-)$ are selected in decreasing order. (For a node with only one child, the sign is irrelevant.) In the simplest scheme (\texttt{GenAltA} in \cite{Ko:1992}), illustrated in Fig. \ref{fig:GenAltA} for $\vec{k}=(2,1,1)$, the nodes are assigned signs that
\emph{alternate} across each level, starting with $+$ on the left. For example, at level 1 (just below the root), the signs from left to right are $+, -, +$. The resulting permutations (reading the bold permutations across the bottom from left to right) indeed have the single-swap (Gray) property \eqref{Grayproperty}. Optimized versions of this algorithm are also presented in \cite{Ko:1992}.

\section{Noise and error detection}\label{sec:noise}

We briefly consider here the effect of noise on the fidelity of our state-preparation algorithm, assuming a simple noise model. We illustrate the improvement in fidelity achieved by the post-selection on the expected weight $\vec{k}$ that is noted at the end of Sec. \ref{sec:full}.

\subsection{Depolarizing noise model}\label{sec:noise-model}

We consider the single-qudit depolarizing channel of strength $p \in [0,1]$, which is defined by \cite{Nielsen:2019}
\begin{equation}
\mathcal{E}(\rho) = (1-p)\,\rho + p\,\frac{\mathbb{I}}{d} \,.
\label{noise-channel}
\end{equation}
In order to implement this channel in cirq, we formulate it in terms of Kraus operators $K_\alpha$ such that 
\begin{equation}
\mathcal{E}(\rho) =\sum_\alpha K_\alpha\, \rho\, K_\alpha^\dagger\,, \qquad 
\sum_\alpha K_\alpha^\dagger\, K_\alpha = \mathbb{I} \,.
\end{equation}
To this end, we define the generalized Pauli operators
\begin{equation}
W_{ab} = {\rm X}^a\, {\rm Z}^b \,, \qquad a\,, b = 0, 1, \ldots, d-1 \,,
\end{equation}
where ${\rm X}$ is the shift operator \eqref{Xgate}, and ${\rm Z}$ is defined by
\begin{equation}
{\rm Z}\, |\mu\rangle = \omega^\mu|\mu \rangle\,, \qquad 
 \omega = e^{\frac{2\pi i}{d}} \,, \qquad \mu = 0, 1, \ldots, d-1 \,.
\label{Zgate}
\end{equation}
Using the fact that averaging any density matrix $\rho$ (for which $\tr \rho =1$) over these operators yields the maximally mixed state
\begin{equation}
\frac{1}{d^2}\sum_{a,b=0}^{d-1} W_{ab}\,\rho\,W_{ab}^\dagger = \frac{\mathbb{I}}{d} \,, 
\end{equation}
one can show that the needed Kraus operators are given by
\begin{equation}
K_{00} =\sqrt{1-p+\frac{p}{d^2}}\mathbb{I}\,, \qquad 
K_{ab} = \sqrt{\frac{p}{d^2}} W_{ab}\quad \text{for }(a,b)\neq(0,0) \,.
\end{equation}

For simplicity, we assume that this noise is applied to each qudit in the circuit that is affected by a gate; hence, $p$ is the per-qudit depolarizing probability applied to each participating qudit in each time-step (layer) of the circuit.

\subsection{Raw and post-selected fidelity}\label{sec:noise-results}

We use \cite{GitHubRN} cirq's density matrix simulator with this noise model to simulate the preparation of $SU(d)$ Dicke states (see Sec. \ref{sec:Dicke}). The ``raw'' fidelity of the noisy state $\rho$ and the pure target state $|\psi\rangle = |D^n(\vec{k})\rangle$ is given by
\begin{equation}
F_{\rm{raw}} = \langle \psi |\rho|\psi \rangle \,,
\end{equation}
which is the square of the Nielsen–Chuang fidelity, Eq. (9.60) of Ref. \cite{Nielsen:2019}.
Post-selection on the expected weight $\vec{k}$ corresponds to projection on the weight-$\vec{k}$ subspace with the operator
\begin{equation}
P = \sum_{w \in \mathfrak{S}_{M(\vec k)}} | w \rangle \langle w|\,.
\label{projector}
\end{equation}
The probability that $\rho$ is in the weight-$\vec{k}$ subspace is therefore
\begin{equation}
p_{\rm success} = \tr( P \rho) \,,
\end{equation}
and the post-selected fidelity is given by
\begin{equation}
F_{\rm{post}} = \frac{\langle\psi|P\rho P|\psi\rangle}{\tr(P\rho)} = \frac{F_{\rm raw}}{p_{\rm success}} \,,
\end{equation}
the last equality because the target state already lies in the subspace ($P|\psi\rangle=|\psi\rangle$). 

\begin{figure}[htb]
	\centering
	\begin{subfigure}{0.49\textwidth}
      \centering
\includegraphics[width=0.9\linewidth]{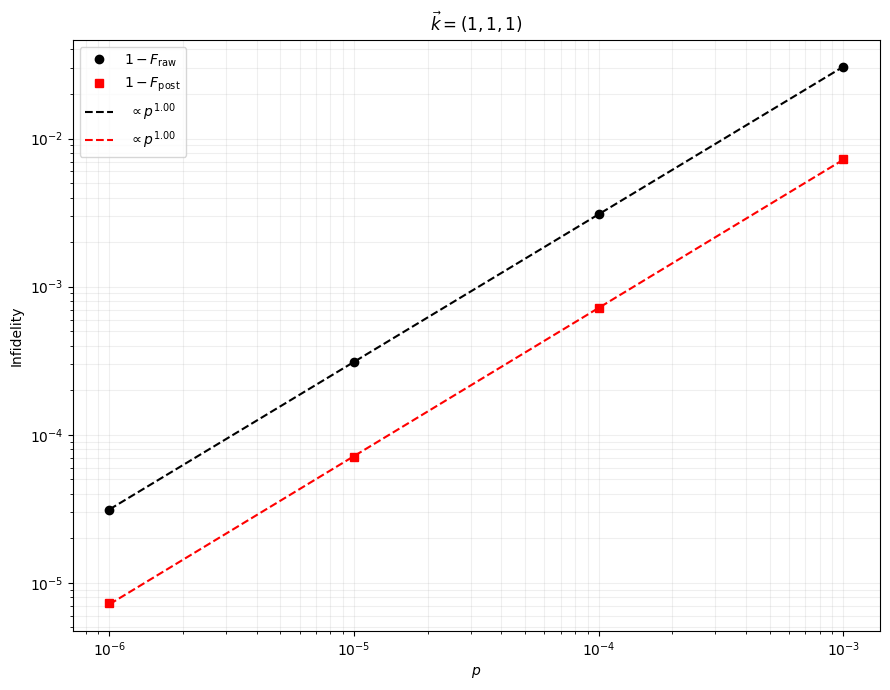}
\caption{The case $\vec{k}=(1,1,1)$}
\label{fig:k111}
    \end{subfigure}%
    \begin{subfigure}{0.49\textwidth}
        \centering
\includegraphics[width=0.9\linewidth]{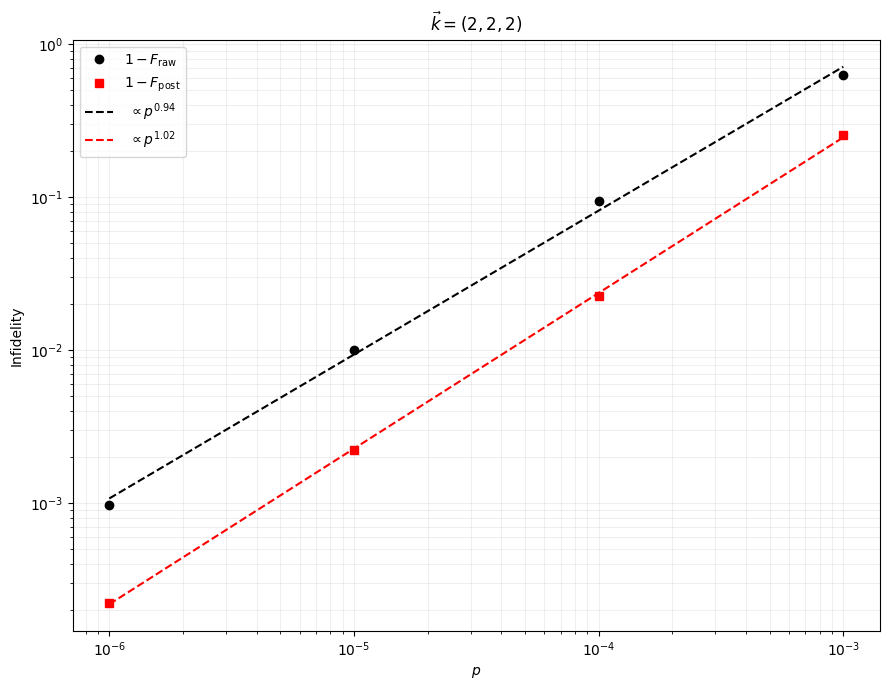}
\caption{The case $\vec{k}=(2,2,2)$}
\label{fig:k222}
	 \end{subfigure}	
\caption{Infidelities $1-F_{\rm raw}$ and $1-F_{\rm{post}}$ as functions of $p$, for
preparing Dicke states $|D^n(\vec{k})\rangle$
for two representative values of $\vec{k}$}
\label{fig:noise}
\end{figure}

Plots of the raw and post-selected
infidelities as functions of $p$, for two representative values of $\vec{k}$, are shown in Fig. \ref{fig:noise}.
The linear behavior $1-F \propto p$ is the expected signature of independent gate errors accumulating linearly, cf. \eqref{fidelity}. Note that the post-selected infidelity (in red) is a factor of a few smaller than the raw infidelity (in black).

\section{Descendant states}\label{sec:descendants}

The on-shell Bethe states \eqref{BA1} are $SU(3)$ highest-weight states that are annihilated by the raising operators \cite{Liashyk:2026dhm}
\begin{equation}
E^{01}\, |\{u \}; \{v \}
\rangle = 0 \,, \qquad
E^{02}\, |\{u \}; \{v \}
\rangle = 0 \,, \qquad
E^{12}\, |\{u \}; \{v \}
\rangle = 0 \,, 
\label{su3hw}
\end{equation}
where
\begin{equation}
E^{ab} = \sum_{i=1}^{n} e^{ab}_i \,, \qquad e^{ab} = |a\rangle \langle b| \,.
\end{equation}
Together with their lower-weight descendants, obtained by acting on the highest-weight states with lowering operators $E^{10}, E^{20}, E^{21}$, they form $SU(3)$ irreducible representations with the common energy \eqref{energyBA}.

We consider here the problem of preparing such descendant states on a quantum computer. Directly applying lowering operators on Bethe states is not an option, since the lowering operators are not unitary. Moreover, introducing additional infinite Bethe roots is also not an option, since the corresponding amplitudes \eqref{BA2} then become indeterminate. As a warm-up, we first treat the simpler $SU(2)$ case in Sec. \ref{sec:SU2descendants} before moving on to the $SU(3)$ case in Sec. \ref{sec:SU3descendants}. We shall see that the descendant states can be prepared similarly to the Bethe states, except with different amplitudes, see \eqref{su3Descendants}-\eqref{su3DescendantsF}.

\subsection{$SU(2)$ descendants}\label{sec:SU2descendants}

For the $SU(2)$-invariant spin-1/2 Heisenberg Hamiltonian of length $n$ with periodic boundary conditions, let us write an (unnormalized) on-shell Bethe state with $m$ Bethe roots as
\begin{equation}
|B_m\rangle = \sum_{\substack{Y \subset \{1, \ldots, n\}\\
|Y|=m}} f(Y) |Y\rangle \,,
\label{su2BetheState}
\end{equation}
where the sum is over all subsets $Y$ of $\{1, \ldots, n\}$ with cardinality $|Y|= m \in [0, \frac{n}{2}]$; $|Y\rangle$ is the $n$-qubit computational basis state with 1's at positions $Y$ and 0's elsewhere; and $f(Y)$ is the
coordinate Bethe ansatz wavefunction, which depends on the $m$ Bethe roots.
These states are $SU(2)$ highest-weight states \cite{Faddeev:1996iy}
\begin{equation}
S^+\, |B_m\rangle = 0 \,,  \qquad S^+ = E^{01} \,,
\end{equation}
with $s=s^z =\frac{n}{2}-m$, where $s$ is the spin and $s^z$ is the eigenvalue of $S^z = \sum_{i=1}^n \frac{1}{2}\sigma_i^z$.

The $SU(2)$ descendant states are obtained by repeatedly applying the lowering operator $S^- = E^{10}$ to the Bethe state. Observe that\footnote{Here and below we do not pay attention to overall factors, as indicated with the proportionality symbol $\propto$.}
\begin{equation}
\left(S^- \right)^\ell \, |Y\rangle \propto \sum_{\substack{Z \subset \{1, \ldots, n\}\\
|Z|=\ell\,, \ Z \cap Y = \emptyset}}  |Y \cup Z \rangle 
= \sum_{\substack{X \subset \{1, \ldots, n\}\\
|X|=\bar{m}\,, \ X \supset Y}}  |X \rangle \,,
\label{observe}
\end{equation}
where $X:=Y \cup Z$ is the set of positions of 1's, whose cardinality is
$\bar{m} := m + \ell$. Applying $\left(S^- \right)^\ell$ to the Bethe state \eqref{su2BetheState}, we obtain
\begin{align}
\left(S^- \right)^\ell \,  |B_m\rangle &=  \sum_{\substack{Y \subset \{1, \ldots, n\}\\
|Y|=m}} f(Y)\, \left(S^- \right)^\ell |Y\rangle \,, \nonumber \\ 
&\propto  \sum_{\substack{Y \subset \{1, \ldots, n\}\\
|Y|=m}} f(Y)\, \sum_{\substack{X \subset \{1, \ldots, n\}\\
|X|=\bar{m}\,, \ X \supset Y}}  |X \rangle \,, \nonumber \\ 
&= \sum_{\substack{X \subset \{1, \ldots, n\}\\
|X|=\bar{m}}} \left(  \sum_{\substack{Y \subset X\\ |Y|=m}} f(Y) \right) |X\rangle \,,
\end{align}
where the result \eqref{observe} has been used in the second line, and
the order of summations has been interchanged in the last line. We conclude that the $SU(2)$ descendant states are given by
\begin{equation}
\left(S^- \right)^\ell \,  |B_m\rangle  \propto 
\sum_{\substack{X \subset \{1, \ldots, n\}\\
|X|= \bar{m}}} F(X) |X\rangle \,, \qquad 
F(X) = \sum_{\substack{Y \subset X\\ |Y|=m}} f(Y) \,,
\label{su2Descedants}
\end{equation}
which can be prepared on a quantum computer similarly to the Bethe state \eqref{su2BetheState} itself \cite{Raveh:2024llj, Farias:2024ejh}.

As a simple example of the result \eqref{su2Descedants}, let us consider the case $m=0$, for which the Bethe state is the reference state $|B_0\rangle = |0\rangle^{\otimes n}$, with $Y=\emptyset$ and $f(Y)=1$. It follows from  \eqref{su2Descedants} that $F(X)=1$, hence
\begin{equation}
\left(S^- \right)^\ell \, |0\rangle^{\otimes n} \propto 
\sum_{\substack{X \subset \{1, \ldots, n\}\\
|X|= \ell}} |X\rangle \,,
\end{equation}
which we recognize as ordinary $SU(2)$ Dicke states with Hamming weight $\ell$. 

\subsection{$SU(3)$ descendants}\label{sec:SU3descendants}

Let us now rewrite the unnormalized Bethe states \eqref{BA1}, analogously to \eqref{su2BetheState},  as
\begin{equation}
|B_{m_1, m_2}\rangle = \sum_{\substack{Y \subset \{1, \ldots, n\}\\
|Y|=m_1}} \sum_{\substack{Y_2 \subset Y\\ |Y_2|=m_2}} f(Y_2 \subset Y)\, 
|Y_2 \subset Y\rangle \,,
\label{su3BetheState}
\end{equation}
where $|Y_2 \subset Y\rangle$ denotes the $n$-qutrit computational basis state with 2's at positions $Y_2$, 1's at positions $Y\backslash Y_2$, and 0's elsewhere; and $f(Y_2 \subset Y)$ corresponds to the amplitude \eqref{BA2}. For these basis states, the
number of 2's is $m_2$, and the number of 1's is $m_1-m_2$. As already noted, $|B_{m_1, m_2}\rangle$ is an $SU(3)$ highest-weight state \eqref{su3hw}.

The $SU(3)$ descendant states of the Bethe state \eqref{su3BetheState}
can be obtained in terms of the Poincar\'e-Birkhoff-Witt (PBW) basis \cite{Hall:2015} as
\begin{equation}
\left( E^{21} \right)^a\, \left( E^{20} \right)^b\,  
\left( E^{10} \right)^c\, |B_{m_1, m_2}\rangle \,.
\label{PBW}
\end{equation}
Compared with $|B_{m_1, m_2}\rangle$,
these states have $\ell_2$ additional 2's, and $\ell_1 - \ell_2$ additional 1's, where
\begin{equation}
    \ell_1 :=  b + c\,, \qquad \ell_2 := a + b \,.
\label{ell1ell2}
\end{equation}
Hence, for the descendant state \eqref{PBW},
the total number of 2's is $\bar{m}_2 := m_2 + \ell_2 = m_2 + a + b$; and the total number of 1's is $\bar{m}_1-\bar{m}_2$, where $\bar{m}_1 := m_1 + \ell_1 = m_1 + b + c$.

\begin{figure}[htbp]
    \centering
    \begin{adjustbox}{width=0.35\textwidth}
\begin{tikzpicture}
    \draw[fill=white, thick] (-4.5,-4) rectangle (4.5,4);
    \node[anchor=north west] at (-4.3, 3.8) {\textbf{Lattice} $\{1, \dots, n\}$};
    
    \fill[cyan!40] (-3, -0.5) rectangle (0, 2.5); 
    \fill[cyan!40] (0, -0.5) rectangle (3, 2.5);  
    
    \draw[thick] (-3, -0.5) rectangle (0, 2.5);
    \draw[thick] (0, -0.5) rectangle (3, 2.5);
    
    \fill[gray!50] (-1, 0) rectangle (1, 2);
    
    \draw[thick, dashed] (-3, -0.5) rectangle (0, 2.5);
    \draw[thick, dashed] (0, -0.5) rectangle (3, 2.5);

    \draw[thick] (-1, 0) rectangle (1, 2);
    
    \filldraw[fill=gray!50, draw=black, thick] (-2.6, 0) rectangle (-1.4, 2);  
    
    \filldraw[fill=gray!50, draw=black, thick] (-1, -3) rectangle (1, -1.5); 

    \node at (-1.5, 2.8) {$Y$};
    \node at (-2.0, 1) {$Y_2$};
    \node at (0, 1) {$Z_2$};
    \node at (1.5, 2.8) {$Z_1$};
    \node at (0, -2.25) {$Z$};

    \node[anchor=south west] at (-4.3, -3.8) {\textbf{State Key:}};
    \draw[fill=gray!50] (-2.3, -3.7) rectangle (-1.9, -3.3) node[pos=.5] {2};
    \draw[fill=cyan!40] (-1.4, -3.7) rectangle (-1.0, -3.3) node[pos=.5] {1};
    \draw[fill=white] (-0.5, -3.7) rectangle (-0.1, -3.3) node[pos=.5] {0};
\end{tikzpicture}
\end{adjustbox}
\caption{Venn diagram representing subsets of $\{1, \ldots, n\}$ entering into the derivation of \eqref{observe2}. The shadings indicate the states (0, 1 or 2) corresponding to these subsets. Not shown: $X_2=Y_2 \cup Z \cup Z_2$ and $X=Y \cup Z_1 \cup Z$.}
    \label{fig:sets}
\end{figure}
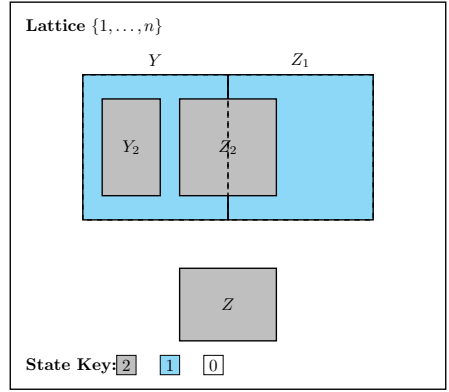

Let us apply the lowering operators in \eqref{PBW}
on $|Y_2 \subset Y\rangle$ step-by-step:
\begin{equation}
\left( E^{10} \right)^c\, |Y_2 \subset Y\rangle 
\propto \sum_{\substack{Z_1 \subset \{1, \ldots, n\}\backslash Y\\
|Z_1|=c\,, \ Z_1 \cap Y = \emptyset}}  |Y_2 \subset (Y \cup Z_1) \rangle \,, 
\end{equation}

\begin{align}
\left( E^{20} \right)^b\,  
\left( E^{10} \right)^c\, |Y_2 \subset Y\rangle 
&\propto \sum_{\substack{Z_1 \subset \{1, \ldots, n\}\backslash Y\\
|Z_1|=c\,, \ Z_1 \cap Y = \emptyset}} 
\left( E^{20} \right)^b\, |Y_2 \subset (Y \cup Z_1) \rangle \nonumber \\
&\propto \sum_{\substack{Z_1 \subset \{1, \ldots, n\}\backslash Y\\
|Z_1|=c\,, \ Z_1 \cap Y = \emptyset}} \quad
\sum_{\substack{Z \subset \{1, \ldots, n\}\backslash (Y\cup Z_1)\\
|Z|=b\,, \ Z \cap (Y \cup Z_1) = \emptyset}}
|(Y_2 \cup Z) \subset (Y \cup Z_1 \cup Z) \rangle \,,
\end{align}

\begin{align}
&\left( E^{21} \right)^a\, \left( E^{20} \right)^b\,  
\left( E^{10} \right)^c\, |Y_2 \subset Y\rangle \nonumber \\
&\propto \sum_{\substack{Z_1 \subset \{1, \ldots, n\}\backslash Y\\
|Z_1|=c\,, \ Z_1 \cap Y = \emptyset}} \quad
\sum_{\substack{Z \subset \{1, \ldots, n\}\backslash (Y\cup Z_1)\\
|Z|=b\,, \ Z \cap (Y \cup Z_1) = \emptyset}}
\left( E^{21} \right)^a\, |(Y_2 \cup Z) \subset (Y \cup Z_1 \cup Z) \rangle \nonumber \\
&\propto \sum_{\substack{Z_1 \subset \{1, \ldots, n\}\backslash Y\\
|Z_1|=c\,, \ Z_1 \cap Y = \emptyset}} \quad
\sum_{\substack{Z \subset \{1, \ldots, n\}\backslash (Y\cup Z_1)\\
|Z|=b\,, \ Z \cap (Y \cup Z_1) = \emptyset}} \quad
\sum_{\substack{Z_2 \subset ((Y\backslash Y_2) \cup Z_1 ) \\
|Z_2|=a}}
|(Y_2 \cup Z \cup Z_2) \subset (Y \cup Z_1 \cup Z) \rangle 
\nonumber \\
&= \sum_{\substack{Z_1 \subset \{1, \ldots, n\}\backslash Y\\
|Z_1|=c\,, \ Z_1 \cap Y = \emptyset}} \quad
\sum_{\substack{Z \subset \{1, \ldots, n\}\backslash (Y\cup Z_1)\\
|Z|=b\,, \ Z \cap (Y \cup Z_1) = \emptyset}} \quad
\sum_{\substack{Z_2 \subset ((Y\backslash Y_2) \cup Z_1 ) \\
|Z_2|=a}}
|X_2 \subset X \rangle \,,
\label{observe2}
\end{align}
where $X_2:=Y_2 \cup Z \cup Z_2$ and $X:=Y \cup Z_1 \cup Z$, whose cardinalities are $\bar{m}_2$ and $\bar{m}_1$ (defined below \eqref{ell1ell2}), respectively. Note that $X_2$ is the set of positions of 2's, and $X\backslash X_2$ is the set of positions of 1's. 
It is helpful to visualize the various subsets in a Venn-type diagram as in 
Fig. \ref{fig:sets}.

Recalling \eqref{su3BetheState}, we obtain
\begin{align}
&\left( E^{21} \right)^a\, \left( E^{20} \right)^b\,  
\left( E^{10} \right)^c\, |B_{m_1, m_2}\rangle \nonumber \\
&= \sum_{\substack{Y \subset \{1, \ldots, n\}\\
|Y|=m_1}} \sum_{\substack{Y_2 \subset Y\\ |Y_2|=m_2}} f(Y_2 \subset Y)\, 
\left( E^{21} \right)^a\, \left( E^{20} \right)^b\,  
\left( E^{10} \right)^c\, |Y_2 \subset Y\rangle  \nonumber \\
&\propto \sum_{\substack{Y \subset \{1, \ldots, n\}\\
|Y|=m_1}} 
\sum_{\substack{Y_2 \subset Y\\ |Y_2|=m_2}} f(Y_2 \subset Y)\,
\sum_{\substack{Z_1 \subset \{1, \ldots, n\}\backslash Y\\
|Z_1|=c\,, \ Z_1 \cap Y = \emptyset}} \quad
\sum_{\substack{Z \subset \{1, \ldots, n\}\backslash (Y\cup Z_1)\\
|Z|=b\,, \ Z \cap (Y \cup Z_1) = \emptyset}} \quad
\sum_{\substack{Z_2 \subset ((Y\backslash Y_2) \cup Z_1 ) \\
|Z_2|=a}}
|X_2 \subset X \rangle \nonumber \\
&= \sum_{\substack{X \subset \{1, \ldots, n\}\\ |X|=\bar{m}_1}}
\sum_{\substack{X_2 \subset X\\ |X_2|=\bar{m}_2}}
\left(
\sum_{\substack{Z \subset X_2 \\ |Z|=b}}
\sum_{\substack{Y_2 \subset (X_2\backslash Z) \\ |Y_2|=m_2}}
\sum_{\substack{Y \subset (X\backslash Z) \\ |Y|=m_1\,, \ Y \supset Y_2}} f(Y_2 \subset Y)
\right) |X_2 \subset X \rangle \,,
\end{align}
where the result \eqref{observe2} has been used in the second step, and
the order and structure of summations have been changed in the last step.
(For given sets $X$ and $X_2 \subset X$, we first fix $Z \subset X_2$, which has no elements in common with any of the sets $Y, Y_2, Z_1, Z_2$. The fact $X_2\backslash Z = Y_2 \cup Z_2$ implies that $Y_2 \subset (X_2\backslash Z)$; and the fact $X\backslash Z = Y \cup Z_1$ implies that $Y \subset (X\backslash Z)$. Once $Y_2$ and $Y$ are fixed, then $Z_1$ and $Z_2$ are also fixed, since $Z_1 = X\backslash (Z \cup Y)$ and $Z_2 = X_2\backslash (Z \cup Y_2)$, see Fig. \ref{fig:sets}.)
We conclude that the $SU(3)$ descendant states are given by
\begin{equation}
\left( E^{21} \right)^a\, \left( E^{20} \right)^b\,  
\left( E^{10} \right)^c\, |B_{m_1, m_2}\rangle 
\propto \sum_{\substack{X \subset \{1, \ldots, n\}\\ |X|=\bar{m}_1}}
\sum_{\substack{X_2 \subset X\\ |X_2|=\bar{m}_2}}
F(X_2 \subset X)\,
|X_2 \subset X \rangle \,, 
\label{su3Descendants}
\end{equation}
where
\begin{equation}
F(X_2 \subset X) = \sum_{\substack{Z \subset X_2 \\ |Z|=b}}
\sum_{\substack{Y_2 \subset (X_2\backslash Z) \\ |Y_2|=m_2}}
\sum_{\substack{Y \subset (X\backslash Z) \\ |Y|=m_1\,, \ Y \supset Y_2}} 
f(Y_2 \subset Y)  \,.
\label{su3DescendantsF}
\end{equation}
The states \eqref{su3Descendants} can be prepared on a quantum computer similarly to the Bethe state \eqref{su3BetheState} itself, except now with the amplitudes $F(X_2 \subset X)$ \eqref{su3DescendantsF} that (like the amplitudes $f(Y_2 \subset Y)$ of the Bethe state) are to be computed classically. In principle, all eigenstates of the Hamiltonian \eqref{HeisHam} corresponding to regular (non-singular) solutions of the Bethe equations \eqref{BE1}-\eqref{BE2} can be prepared in this way. 

As a simple example of the result \eqref{su3Descendants}, let us consider the case $m_1=m_2=0$, for which the Bethe state is the reference state $|B_{0,0}\rangle = |0\rangle^{\otimes n}$, with $Y_2=Y=\emptyset$ and $f(Y_2 \subset Y)=1$. All of the descendant states of $|B_{0,0}\rangle$ can be obtained with $b=0$, for which case 
\eqref{su3DescendantsF} implies that $F(X_2 \subset X)=1$, and we obtain
\begin{equation}
\left( E^{21} \right)^a\, \left( E^{10} \right)^c\,  |0\rangle^{\otimes n} \propto 
\sum_{\substack{X \subset \{1, \ldots, n\}\\ |X|= c}} 
\sum_{\substack{X_2 \subset X\\ |X_2|=a}}
|X_2 \subset X \rangle \,,
\end{equation}
which we recognize as $SU(3)$ Dicke states (see Sec. \ref{sec:Dicke})
with weight $\vec{k}=(n-c,c-a,a)$.



\providecommand{\href}[2]{#2}\begingroup\raggedright\endgroup

\end{document}